\documentclass[aps,prd,superscriptaddress,preprint,amsmath,amssymb]{revtex4}
\usepackage{graphicx}
\usepackage{float}
\begin{document}

\title{Probing the anomalous electromagnetic dipole moments of the top quark in $\gamma p$ collision at the LHC, the HL-LHC and the HE-LHC}

\author{M. Koksal}
\email[]{mkoksal@cumhuriyet.edu.tr} \affiliation{Department of
Optical Engineering, Sivas Cumhuriyet University, 58140, Sivas, Turkey}

\begin{abstract}
The determination of the electromagnetic dipole moments of the top quark is one of the most important goals of the top quark physics program in the collider experiments. For this reason, the top quark pair production to investigate the sensitivity on the electric and magnetic dipole moments of the top quark via the process $pp\rightarrow p \gamma p \rightarrow p t \bar{t} X$ at the Large Hadron Collider (LHC), the High-Luminosity Large Hadron Collider (HL-LHC) and High-Energy Large Hadron Collider (HE-LHC) is discussed. We apply pure leptonic and semileptonic decays for top quark pair production in the final state. Moreover, we consider systematic uncertainties of $0,3\%$ and $5\%$. The best limits obtained from the process $pp\rightarrow p \gamma p \rightarrow p t \bar{t} X$ on the anomalous $a_{A}$ and $a_{V}$ couplings are $|a_{A}|=0.0200$ and $a_{V}=[-0.9959; 0.0003]$. Thus, our results indicate that the process $pp\rightarrow p \gamma p \rightarrow p t \bar{t} X$ is a very good perspective to probe the electric and magnetic dipole moments of the top quark at the HL-LHC and the HE-LHC.
\end{abstract}

\maketitle

\section{Introduction}

The Standard Model (SM) of particle physics has been greatly successful in forecasting a wide range of phenomena. However, with the ultimate discovery of the Higgs boson with approximately 125 GeV mass by CMS and ATLAS Collaborations at the LHC, the SM has obtained a significant achievement \cite{1,2}. On the other hand, this model leaves some questions unanswered such as neutrino oscillations, the strong CP problem and matter-antimatter asymmetry, etc.
Thus, it is thought to be embedded in a more fundamental theory where its effects can be observed at higher energy scales.

Among the all observed elementary particles of the SM, the largest mass particle is the top quark with a mass of $173.0\pm 0.4$ GeV \cite{3}. Investigation of the interactions of the top quark is important not only for the dynamics of electroweak symmetry breaking but also for testing of SM and new physics beyond SM. Up to now, this heavy quark produced by the various processes at the Tevatron and LHC was examined in detail. In this case, in addition to detecting the top quark, it has been a tremendous motivation to examine the characteristics and potential of the top quark in both decay and production. The complicated experimental results of the LHC are accomplished by precise theoretical predictions within the framework of the SM and beyond the SM. Many of its properties are still poorly constrained such as the electric and magnetic dipole moments and the chromoelectric and chromomagnetic dipole moments. For this reason, important new insights on the properties of the top quark will be one of the tasks of the LHC. Especially, the anomalous $t\bar{t}\gamma$ couplings that can define the electromagnetic dipole moments of the top quark, which is the subject of this study, have been investigated extensively at lepton-lepton, hadron-hadron colliders and lepton-hadron colliders.

One of the significant events in the field of fundamental interactions currently defined by the SM is the violation of CP symmetry. CP violation in the SM is identified with a complex phase in the CKM matrix. However, this information from the CKM matrix for CP violation cannot define the matter-antimatter asymmetry in the universe. This asymmetry is one of the principal questions in the SM. Therefore, the measurement of large amounts of CP violation in
the top quark events in the examined processes can be a proof of new physics beyond the SM. Investigation of new physics beyond the SM, some of
the intrinsic properties of the top quark are examined in the context of its dipole moments such as the magnetic dipole moment arising from one-loop level and the corresponding electric dipole moment that is defined as a source of CP violation coming from the three-loop level in the SM \cite{4,5}.

The value for the magnetic dipole moment of the top quark predicted by the SM is $0.02$. This value can be tested in the current and the upcoming experiments. In addition, the electric dipole moment of the top quark in the SM is suppressed with a value of less than $10^{-30} ({\rm e \hspace{0.8mm} cm})$. Besides,
it is highly attractive for the investigation of new physics beyond the SM. If there is a sign of new physics beyond the SM in the examined processes at the LHC, then the top quark may have an the electric dipole moment higher than the SM value.

In the literature, there have been different proposals to observe the electric and magnetic dipole moments of the top quark. Studies at the Tevatron and the LHC were recommended to obtain the electromagnetic dipole moments of the top quark in measurements of the processes $p \bar{p}\rightarrow t\bar{t}\gamma$ \cite{6}, $p p\rightarrow t j\gamma$ \cite{7,8} and $p p\rightarrow p \gamma^{*} \gamma^{*} p\rightarrow p t \bar{t} p$ \cite{9}. The reactions $e^{-}e^{+}\rightarrow t \bar{t}$ \cite{10}, $\gamma e\rightarrow \bar{t} b \nu_{e}$ \cite{11}, $e^{-}e^{+}\rightarrow e^{-}\gamma^{*} e^{+} \rightarrow \bar{t} b \nu_{e} e^{+}$ \cite{11}, $\gamma \gamma \rightarrow t \bar{t}$ \cite{12} and $e^{-}e^{+}\rightarrow e^{-}\gamma^{*} \gamma^{*} e^{+} \rightarrow e^{-}t \bar{t}e^{+}$ \cite{12} at the future $e^{-}e^{+}$ linear colliders and their operating modes of $e \gamma$, $e \gamma^{*}$, $\gamma \gamma$ and $\gamma^{*} \gamma^{*}$ were investigated to set the limits on the electric and magnetic dipole moments of the top quark. However, the reactions $ep\rightarrow \bar{t} \nu_{e} \gamma$ \cite{13}, $ep\rightarrow e \gamma^{*} p\rightarrow e t \bar{t} X$ \cite{14}, $ep\rightarrow e \gamma^{*} p\rightarrow e t W X$ \cite{14} and $ep\rightarrow e \gamma^{*} \gamma^{*} p\rightarrow e t \bar{t} p$ \cite{15} in phenomenological investigations on the future $ep$ colliders are considered. Finally, Ref. \cite{16} studied the limits on the electromagnetic dipole moments of the top quark that are calculated from measurements of the semi-inclusive decays $b \rightarrow s\gamma$, and of $t \bar{t} \gamma$ production at the Tevatron and the LHC. Also, a complementary way to access the electric dipole moment of the top is through their indirect effects, such as the resulting, radiatively-induced the electric dipole moment of the electron. In summary, all of the current limits on the electric and magnetic dipole moments of the top quark are represented in Table I.

\begin{table}[h]
\caption{Sensitivity limits on the magnetic and electric dipole moments of top quark through different processes at $pp$, $e^{-} e^{+}$ and $ep$ colliders}
\begin{center}
\begin{tabular}{|c|c|c|}
\hline\hline
{\bf Processes}  &    {\bf $a_{V}$}  &    {\bf $a_{A}$}  \\
\hline
$p p\rightarrow t\bar{t}\gamma$ \cite{6}           &   $ (-0.200, 0.200)$ & $(-0.100, 0.100) $ \\
\hline
$p p\rightarrow t j\gamma$ \cite{8}           &   $ (-0.220, 0.210)$ & $ (-0.200, 0.200) $   \\
\hline
$pp \to p\gamma^*\gamma^*p\to pt\bar t p $ \cite{9}      &   $ (-0.4588, 0.0168)$ & $ (-0.0815, 0.0815) $  \\
\hline
$e^+e^- \to t\bar t$ \cite{10}         &   $ (-0.002, 0.002)$ & $ (-0.001, 0.001) $   \\
\hline
$\gamma e\rightarrow \bar{t} b \nu_{e}$ \cite{11}         &   $ (-0.027, 0.036)$ & $ (-0.031, 0.031) $ \\
\hline
$e^{-}e^{+}\rightarrow e^{-}\gamma^{*} e^{+} \rightarrow \bar{t} b \nu_{e} e^{+}$ \cite{11}         &   $ (-0.054, 0.092)$ & $ (-0.071, 0.071) $  \\
\hline
$\gamma \gamma \rightarrow t \bar{t}$ \cite{12}         &   $ (-0.220, 0.002)$ & $ (-0.020, 0.020) $ \\
\hline
$e^{-}e^{+}\rightarrow e^{-}\gamma^{*} \gamma^{*} e^{+} \rightarrow e^{-}t \bar{t}e^{+}$ \cite{12}         &   $ (-0.601, 0.015)$ & $ (-0.089, 0.089) $ \\
\hline
$ep\rightarrow \bar{t} \nu_{e} \gamma$ \cite{13} &   $ (-0.204, 0.185)$ & $ (-0.193, 0.193) $  \\
\hline
$ep\rightarrow e \gamma^{*} p\rightarrow e t W X$ \cite{14} &   $ (-0.204, 0.185)$ & $ (-0.193, 0.193) $    \\
\hline
$ep\rightarrow e \gamma^{*} p\rightarrow \bar{t} \nu_{e} b p$ \cite{14} &   $ (-0.089, 0.085)$ & $ (-0.087, 0.087) $    \\
\hline
$ep\rightarrow e \gamma^{*} \gamma^{*} p\rightarrow e t \bar{t} p$ \cite{15} &   $ ( -0.468, 0.0177)$ & $ (-0.088, 0.088) $    \\
\hline
Radiative $b\to s\gamma$ transitions \cite{16}      &   $ (-2, 0.3)$ & $ (-0.5, 1.5) $  \\
\hline
\end{tabular}
\end{center}
\end{table}

For the present, the LHC has finalized its phase 2 and has closed for an upgrade between 2019 with 2020 years. In later times, it is going to operate at a center-of-mass energy of 14 TeV during the period 2021-2023 and is going to collect almost 300 fb$^{-1}$ of additional data for each detector. However, there will be a major upgrade of the LHC to High-Luminosity LHC (HL-LHC) between 2023 with 2026. Therefore, HL-LHC is anticipated to operate for ten years until 2036. At the end of this duration, it is estimated that each detector will collect approximately 3000 fb$^{-1}$ data. Other colliders other than HL-LHC are also discussed. Also, the High-Energy LHC (HE-LHC) with a center-of-mass energy of 27 TEV at CERN is designed. It will collect a dataset corresponding to an integrated luminosity of 10-15 ab$^{-1}$.

For the new physics research beyond the SM at LHC, $pp$ deep inelastic scattering processes that involve subprocesses of gluon-gluon, quark-quark and quark-gluon collisions are generally investigated in detail. However, due to proton remnants, these processes have not provided very clean environment. Pollution in this environment can occur certain uncertainties and make it tough to observe the signs which may arise from the new physics. Nevertheless, in the literature, exclusive and semi-elastic processes are much less examined. Both of the incoming protons in an exclusive process remains intact and do not dissociate into partons. In addition to this,  only one of the incoming protons in a semi-elastic process dissociates into partons but the other proton remains intact. The exclusive and semi-elastic processes are $\gamma^{*} \gamma^{*}$ and $\gamma^{*} p$, respectively. Among these processes, the cleanest channel is $\gamma^{*} \gamma^{*}$. The exclusive and semi-elastic have simpler final states with respect to $pp$ processes. Therefore, these processes compensate for the advantages of $pp$ processes such as having high center-of-mass energy and high luminosity.

In $\gamma^{*} p$ processes, since one from the incoming protons decomposes into partons they contain more background than $\gamma^{*} \gamma^{*}$ processes.
Besides, $\gamma^{*} p$ processes have effective luminosity and much higher energy compared to $\gamma^{*} \gamma^{*}$ process. This may be significant because of the high energy dependencies of the cross sections containing the new physics parameters. For this reason, $\gamma^{*} p$ processes are anticipated to have a high sensitivity to the anomalous couplings. Photons emitted from one of the proton beams in $\gamma^{*} p$ collision at the LHC can be defined in the framework of the Equivalent Photon Approximation (EPA) \cite{epa,epa1,epa2}. These photons in the EPA have low virtuality. Since protons emit quasi-real photons, they do not decompose into partons. The EPA has many advantages. It aids to obtain crude numerical predictions via easy formulas. In addition to this, the EPA can mainly simplify the experimental analysis because it provides an occasion one to directly get a rough cross-section for $\gamma^{*} \gamma^{*}\rightarrow X$ subprocess via the investigation of the process $pp \rightarrow p X p$. Here, $X$ denotes objects produced in the final state. In the literature, there are a lot of phenomenological studies which are based on the photon-induced processes at the LHC aimed at research for new physics beyond the SM \cite{s1,s2,s3,s4,s5,s6,s7,s8,s9,s10,s11,s12,s13,s14,s15,s16,s17,s18,s19,s20,s21,s22,sa1,sa2,sa3,sa4,sa5,sa6,sa7,sa8,sa9,sa10,sa11}.

\section{Top quark pair production in $\gamma^{*} p$ collisions}

\subsection{The anomalous $t\bar t \gamma$ couplings}

A method for defining possible new physics beyond the SM in a model-independent way is effective Lagrangian approach. This approach is described by high-dimensional operators which cause the anomalous $t\bar{t}\gamma$ coupling. These operators can be defined below \cite{Kamenik,Baur,Aguilar,Aguilar1}

\begin{equation}
{\cal L}_{t\bar t\gamma}=-e Q_t\bar t \Gamma^\mu_{ t\bar t  \gamma} t A_\mu.
\end{equation}

Eq. (1) contains the SM coupling and contributions arising from dimension-six effective operators. Also, $e$ is the proton charge, $Q_t$ shows the top quark electric charge, $A_\mu$ represents the photon gauge field. $\Gamma^\mu_{t\bar t \gamma}$ has the following form

\begin{equation}
\Gamma^\mu_{t\bar t\gamma}= \gamma^\mu + \frac{i}{2m_t}(a_V + i a_A\gamma_5)q_\nu \sigma^{\mu\nu}
\end{equation}

\noindent where $m_t$ is the top quark mass, $q_\nu$ describes the photon four-momentum, $\gamma_5 q_\nu$ term with $\sigma^{\mu\nu}$ breaks the CP symmetry. Thus, $a_A$ parameter describes the strength of a possible CP violation process, which may be caused by new physics beyond the SM. Real $a_V$ and $a_A$ parameters are non-SM couplings and interested in the anomalous magnetic moment and the electric dipole moment of the top quark, respectively. The relations between these parameters and the electromagnetic dipole moments are described as follows

\begin{eqnarray}
a_V&=&Q_t a_t,  \\
a_A&=&\frac{2m_t}{e}d_t.
\end{eqnarray}

\subsection{The cross section of the process $pp\rightarrow p \gamma p \rightarrow p t \bar{t} X$}

A quasi-real photon emitted from one of the two proton beams interacts with the incoming other proton beam, and $\gamma^{*} p$ collisions occur. Symbolic diagram of the process $pp\rightarrow p \gamma^{*} p \rightarrow p t \bar{t} X$ is displayed in Fig. 1.

\begin{figure} [h]
\begin{center}
\includegraphics [width=8cm,height=6cm]{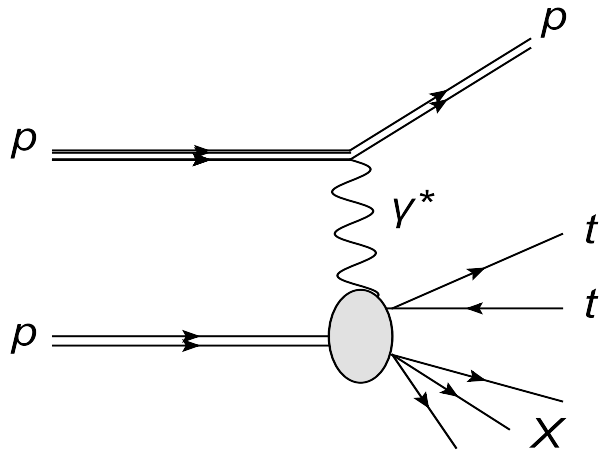}
\caption{Schematic diagram for the process $pp\rightarrow p \gamma^{*} p \rightarrow p t \bar{t} X$ at the LHC.
\label{fig1}}
\end{center}
\end{figure}

The electromagnetic field of the colliding hadrons (protons or heavy ions) at the LHC can be seen as an incoming photon flux, distributed with some density
$dN(\frac{E_{\gamma}}{E}, Q^{2})$. The EPA factorises the dependence on photon virtuality $Q^{2}$ from the cross-section of the photon-induced process to the equivalent photon flux $dN$. If the photon flux originates in a nucleon which is not considered as pointlike, the electric and magnetic form factors should be taken into account.  These factors are defined via the matrix element of the electromagnetic current \cite{gutt}

\begin{eqnarray}
\langle N(P')|j_{\mu}^{em}|N(P)\rangle=\bar{N}(P')[\gamma_{\mu} F_{1}(Q^{2})-i\frac{\sigma_{\mu\nu}q^{\nu}}{2m_{N}}F_{2}(Q^{2})]N(P)
\end{eqnarray}
where $P$ and $P'$ are the 4 - momentum of the nucleon of mass $m_{N}$ before and after photon emission.  $F_{1}$ and $F_{2}$ are the Dirac and Pauli form factors, respectively. $\sigma_{\mu\nu}=\gamma_{\mu}\gamma_{\nu}-\gamma_{\nu}\gamma_{\mu}$, $q=P-P'$.

The Sachs form factors are expressed in terms of $F_{1}$ and $F_{2}$ electromagnetic functions

\begin{eqnarray}
G_{E}(Q^{2})=F_{1}(Q^{2})-\frac{Q^{2}}{4m_{N}}F_{2}(Q^{2}),
\end{eqnarray}

\begin{eqnarray}
G_{M}(Q^{2})=F_{1}(Q^{2})+F_{2}(Q^{2})
\end{eqnarray}

At $Q^{2}=0$, these functions correspond to the total charge and to the magnetic momentum $\mu_{p}$ of the proton, respectively;

\begin{eqnarray}
G_{E}(0)=1,
\end{eqnarray}

\begin{eqnarray}
G_{M}(0)=\mu_{p}=2.79.
\end{eqnarray}

In the usual dipole approximation, the dependence on $Q^{2}$ of the form factors is explicit

\begin{eqnarray}
G_{E}(Q^{2})=\frac{G_{M}(Q^{2})}{\mu_{p}}=(1+\frac{Q^{2}}{Q_{0}^{2}})^{-4},
\end{eqnarray}
where $Q_{0}^{2}=0.71 GeV^{2}$

In the EPA, the photon flux from a proton can then be written in terms of the form factors \cite{bud}:

\begin{eqnarray}
\frac{d N_{\gamma}}{d E_{\gamma} d Q^{2}}=\frac{\alpha}{\pi}\frac{1}{E_{\gamma}Q^{2}}[(1-\frac{E_{\gamma}}{E})(1-\frac{Q_{min}^{2}}{Q^{2}})F_{E}+\frac{E^{2}_{\gamma}}{2E^{2}}F_{M}]
\end{eqnarray}
where
\begin{eqnarray}
Q_{min}^{2}=\frac{m_{p}^{2}E^{2}_{\gamma}}{E(E-E_{\gamma})}.
\end{eqnarray}

$F_{E}$ and $F_{M}$ are functions of the electric and magnetic form factors. These are given below

\begin{eqnarray}
F_{E}=\frac{4 m_{p}^{2}G^{2}_{E}+Q^{2}G_{M}^{2}}{4 m_{p}^{2}+Q^{2}},
\end{eqnarray}

\begin{eqnarray}
F_{M}=G_{M}^{2}.
\end{eqnarray}

Here, the mass of the proton is $m_{p}=0.938$ GeV, $E$ represents the energy of the incoming proton beam.

After integration over $Q^{2}$, equivalent photon spectrum can be given by

\begin{eqnarray}
\frac{d N_{\gamma}}{d E_{\gamma}}=\frac{\alpha}{\pi E_{\gamma}}\{[1-\frac{E_{\gamma}}{E}][\varphi(\frac{Q_{max}^{2}}{Q_{0}^{2}})-\varphi(\frac{Q_{min}^{2}}{Q_{0}^{2}})]
\end{eqnarray}

where the function $\varphi$ is described as follows

\begin{eqnarray}
\varphi(\theta)=&&(1+ay)\left[-\textit{In}(1+\frac{1}{\theta})+\sum_{k=1}^{3}\frac{1}{k(1+\theta)^{k}}\right]+\frac{y(1-b)}{4\theta(1+\theta)^{3}} \nonumber \\
&& +c(1+\frac{y}{4})\left[\textit{In}\left(\frac{1-b+\theta}{1+\theta}\right)+\sum_{k=1}^{3}\frac{b^{k}}{k(1+\theta)^{k}}\right]. \nonumber \\
\end{eqnarray}
Here,

\begin{eqnarray}
y=\frac{E_{\gamma}^{2}}{E(E-E_{\gamma})},
\end{eqnarray}
\begin{eqnarray}
a=\frac{1+\mu_{p}^{2}}{4}+\frac{4m_{p}^{2}}{Q_{0}^{2}}\approx 7.16,
\end{eqnarray}
\begin{eqnarray}
b=1-\frac{4m_{p}^{2}}{Q_{0}^{2}}\approx -3.96,
\end{eqnarray}
\begin{eqnarray}
c=\frac{\mu_{p}^{2}-1}{b^{4}}\approx 0.028.
\end{eqnarray}

The cross section of the process $pp\rightarrow p \gamma^{*} p \rightarrow p t \bar{t} X$ can be calculated by integrating the cross section for the subprocess $\gamma^{*} g \rightarrow t \bar{t}$ over the photon and quark spectra:

\begin{eqnarray}
\sigma(pp\rightarrow p \gamma^{*} p \rightarrow p t \bar{t} X)=\int \int (\frac{d N_{\gamma}}{d x_{1}}) (\frac{d N_{g}}{d x_{2}}) d x_{1} d x_{2} \hat{\sigma}_{\gamma^{*} g \rightarrow t \bar{t}}
\end{eqnarray}
where $x_{1}=\frac{E_{\gamma}}{E}$, $x_{2}$ is the momentum fraction of the proton's momentum carried by the gluon. $\frac{d N_{g}}{d x_{2}}$ is the parton distribution function of the gluon.

As seen in Fig. 2, the reaction $\gamma^{*} g \rightarrow t \bar{t}$ has two Feynman diagrams.

\begin{figure} [h]
\centering
\includegraphics[width=8cm,height=5cm]{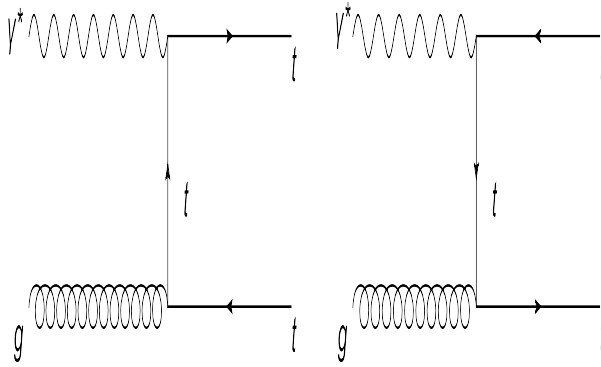}
\caption{Feynman diagrams for the subprocess $\gamma^{*} g \rightarrow t \bar{t}$
\label{fig2}}
\end{figure}

The CalcHEP computer package was used to calculate the cross section of the process $pp\rightarrow p \gamma^{*} p \rightarrow p t \bar{t} X$ including the anomalous $t\bar{t}\gamma$ vertex given in Eq. (2). Thus, we obtain numerically the cross sections as a function of the center-of-mass energies and effective couplings:

- Total cross sections including an anomalous parameter at $\sqrt{s}=14$ TeV:

\begin{eqnarray}
\sigma(a_V)&=&\Bigl[(0.606)a^2_V +(0.658)a_V  +0.481 \Bigr] (pb),   \\
\sigma(a_A)&=&\Bigl[(0.606)a^2_A + 0.481 \Bigr] (pb).
\end{eqnarray}

- Total cross sections including an anomalous parameter at $\sqrt{s}=27$ TeV:

\begin{eqnarray}
\sigma(a_V)&=&\Bigl[(2.091)a^2_V +(2.082) a_V  +1.537 \Bigr] (pb),   \\
\sigma(a_A)&=&\Bigl[(2.091) a^2_A + 1.537 \Bigr] (pb).
\end{eqnarray}

Therefore,

- Total cross section including two anomalous parameters at $\sqrt{s}=14$ TeV:

\begin{eqnarray}
\sigma(a_V, a_A)&=&\Bigl[(0.606)a^2_V +(0.606)a^2_A +(0.658)a_V  +0.481 \Bigr] (pb),
\end{eqnarray}

- Total cross section including two anomalous parameters at $\sqrt{s}=27$ TeV:

\begin{eqnarray}
\sigma(a_V, a_A)&=&\Bigl[(2.091)a^2_V +(2.091)a^2_A +(2.082)a_V  +1.537 \Bigr] (pb).
\end{eqnarray}

In these equations, the independent terms from $a_V$ and $a_A$ parameters indicate the cross section of the SM. In addition, as can be understood from these equations, the linear terms of the anomalous couplings arise from the interference between the anomalous and the SM contribution, whereas the quadratic terms give purely anomalous contribution. Therefore, the total cross sections of the process $pp\rightarrow p \gamma^{*} p \rightarrow p t \bar{t} X$ with respect to the anomalous $a_V$ and $a_A$ couplings are represented in Figs. 3-6.

\begin{figure} [h]
\includegraphics{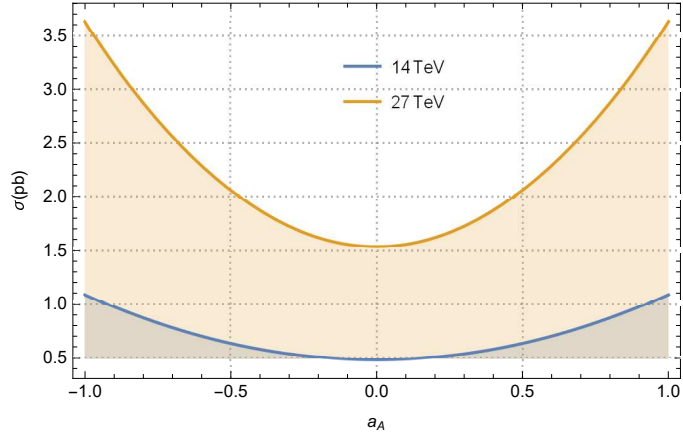}
\caption{The total cross section of the process $pp\rightarrow p \gamma^{*} p \rightarrow p t \bar{t} X$ as a function of the anomalous $a_{A}$ coupling for center-of-mass energies of $\sqrt{s}=14, 27$ TeV.
\label{fig3}}
\end{figure}

\begin{figure} [h]
\includegraphics{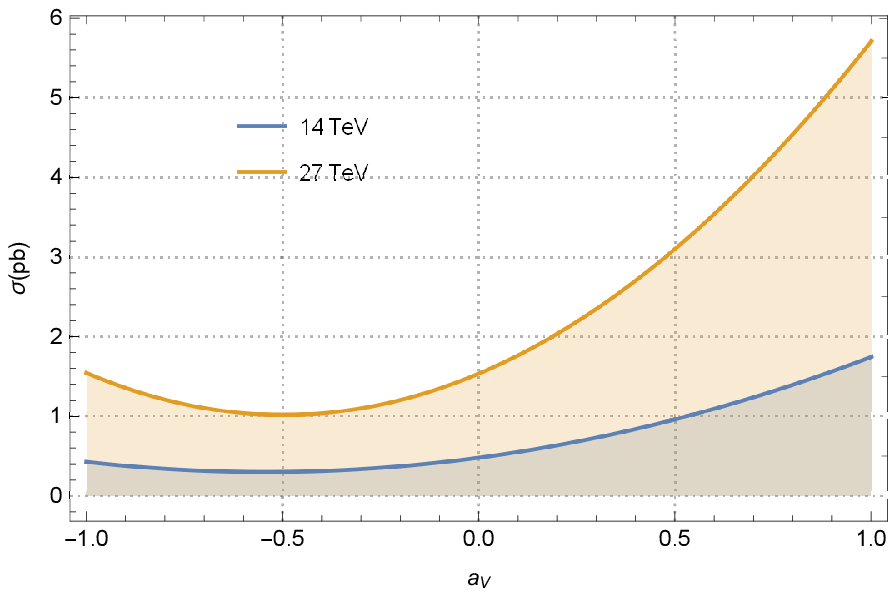}
\caption{The total cross section of the process $pp\rightarrow p \gamma^{*} p \rightarrow p t \bar{t} X$ as a function of the anomalous $a_{V}$ coupling for center-of-mass energies of $\sqrt{s}=14, 27$ TeV.
\label{fig4}}
\end{figure}

\begin{figure} [h]
\includegraphics{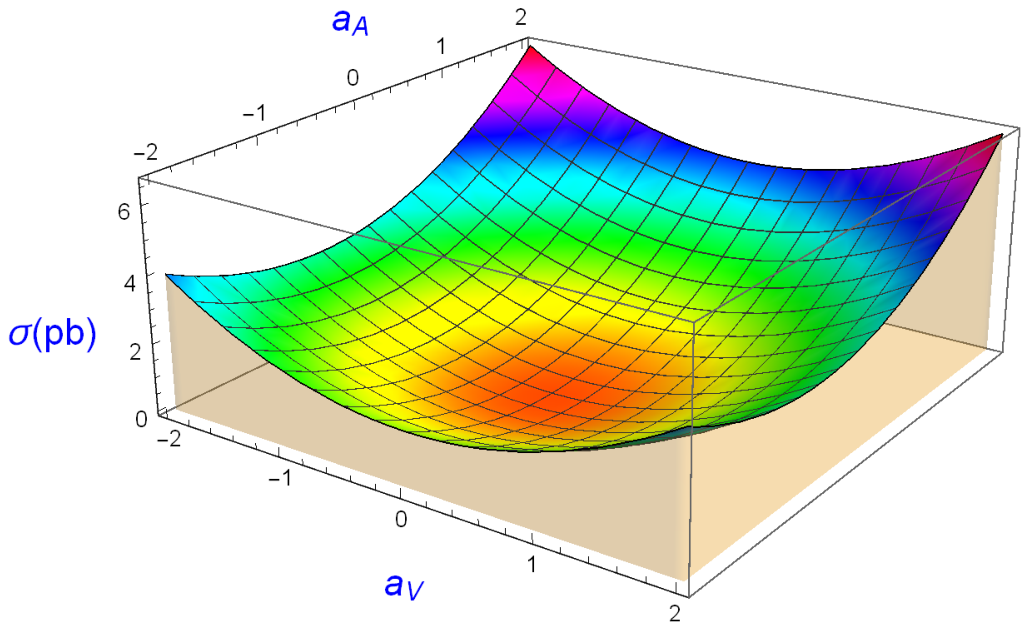}
\caption{The total cross section of the process $pp\rightarrow p \gamma^{*} p \rightarrow p t \bar{t} X$ as a function of the anomalous $a_{A}$ and $a_{V}$ coupling for center-of-mass energy of $\sqrt{s}=14$ TeV.
\label{fig5}}
\end{figure}

\begin{figure} [h]
\includegraphics{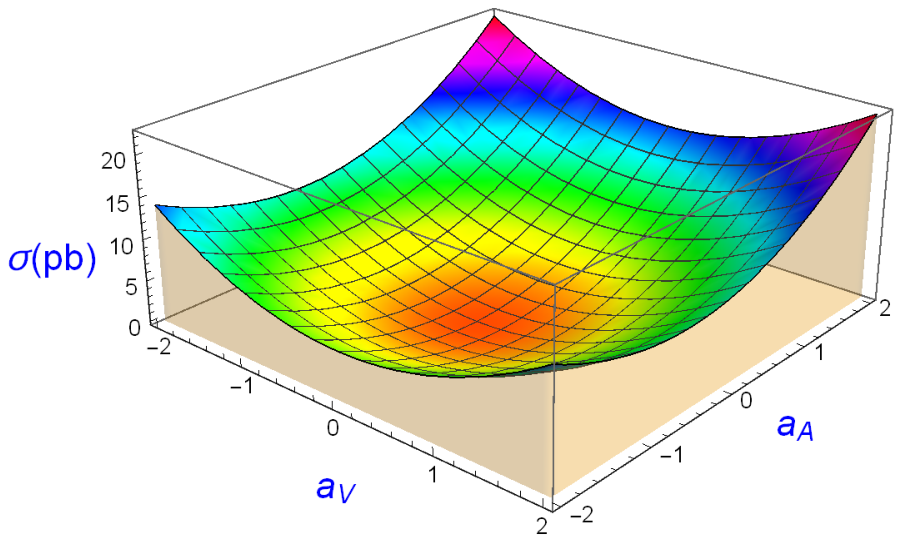}
\caption{The total cross section of the process $pp\rightarrow p \gamma^{*} p \rightarrow p t \bar{t} X$ as a function of the anomalous $a_{A}$ and $a_{V}$ coupling for center-of-mass energy of $\sqrt{s}=27$ TeV.
\label{fig6}}
\end{figure}

In Figs. 3-4, the total cross sections are calculated with considering that only one of the anomalous $a_V$ and $a_A$ couplings have changed while the other coupling is taken into account as zero. As seen from Figs. 3-6, the cross sections of the examined process show a clear dependence on the anomalous $a_V$ and $a_A$ couplings. From the above equations we understand that the anomalous parameters have different CP properties. The cross sections have even powers of the anomalous $a_A$ coupling and a nonzero value of $a_A$ coupling permits a constructive effect on the total cross section. On the other hand, the cross sections contain only odd powers of $a_V$ coupling. In Fig. 4, there are small intervals around $a_V$ in which the cross section that includes new physics beyond the SM is smaller than the SM cross section. Thus, $a_V$ coupling has a partially destructive effect on the total cross section. Fig. 4 represents that the deviation from the SM of the positive part of $a_V$ coupling is greater than the deviation of the negative part. So we expect the sensitivity of the positive part of $a_V$ coupling to be higher than the negative part.

\section{LIMITS ON THE TOP QUARK'S ELECTRIC AND MAGNETIC DIPOLE MOMENTS AT THE LHC, HL-LHC AND HE-LHC}

To obtain the sensitivity on the anomalous couplings, we consider $\chi^{2}$ analysis with a systematic error

\begin{eqnarray}
\chi^{2}=\left(\frac{\sigma_{SM}-\sigma_{NP}(a_{A},a_{V})}{\sigma_{SM}\delta}\right)^{2},
\end{eqnarray}
where $\sigma_{SM}$ is the SM cross section, $\sigma_{NP}(a_{A},a_{V})$ is the total cross section containing contributions from the SM and new
physics, $\delta=\frac{1}{\sqrt{\delta_{stat}^{2}+\delta_{sys}^{2}}}, \delta_{stat}=\frac{1}{\sqrt{N_{SM}}}$ is the statistical error, $N_{SM}=L_{int}\times BR \times \sigma_{SM}\times b_{tag}\times b_{tag}$; $L_{int}$ is the integrated luminosity and $b_{tag}$ tagging efficiency is 0.8. The top quark decays nearly $100\%$ to $b$ quark and $W$ boson. For top quark pair production, we can categorize decay products according to the decomposition of $W$ boson. In our calculations, we consider pure leptonic and semileptonic decays of $W$ bosons in the final state. Thus, while branching ratios for pure leptonic decays of $W$ bosons are BR = 0.123, for semileptonic decays are BR = 0.228.

The inclusive $t\bar{t}$ production cross section using 3.2 fb$^{-1}$ of $\sqrt{s}=$ 13 TeV $pp$ collisions by the ATLAS detector at the LHC is measured \cite{top}. The four uncertainties giving a total relative uncertainty of $4.4\%$ have calculated in the process of determining the cross section of top pair production. These are experimental and theoretical systematic effects, the integrated luminosity and the LHC beam energy. In order to examine the limits on the electromagnetic dipole moments of the top quark, there are also theoretical studies that take into account systematic uncertainties.  The processes $\gamma \gamma \rightarrow t \bar{t}$ and $e^{-}e^{+}\rightarrow e^{-}\gamma^{*} \gamma^{*} e^{+} \rightarrow e^{-}t \bar{t}e^{+}$  with systematic uncertainties of $0, 5, 10\%$ are discussed in Ref. \cite{12}. In Ref. \cite{14}, the processes $\gamma e\rightarrow \bar{t} b \nu_{e}$, $e^{-}e^{+}\rightarrow e^{-}\gamma^{*} e^{+} \rightarrow \bar{t} b \nu_{e} e^{+}$, $ep\rightarrow e \gamma^{*} p\rightarrow \bar{t} \nu_{e} b p$ are studied from $0\%$ to $5\%$ with systematic uncertainties. In Ref. \cite{16}, a $10\%$ total uncertainty for measurements of the process $\gamma e \rightarrow t \bar{t}$ is considered.  In the light of these discussions, systematic error values of $0, 3, 5\%$ are assumed during statistical analysis.

Figs. 7-8 indicate limit values obtained the anomalous $a_{A}$ and $a_{V}$ at $95\%$ C.L. through the process $pp\rightarrow p \gamma p \rightarrow p t \bar{t} X$ at the LHC, the HL-LHC and the HE-LHC. We can easily compare the limits obtained from the LHC, the HL-LHC and the HE-LHC for various integrated luminosities.

\begin{figure} [!h]
\includegraphics{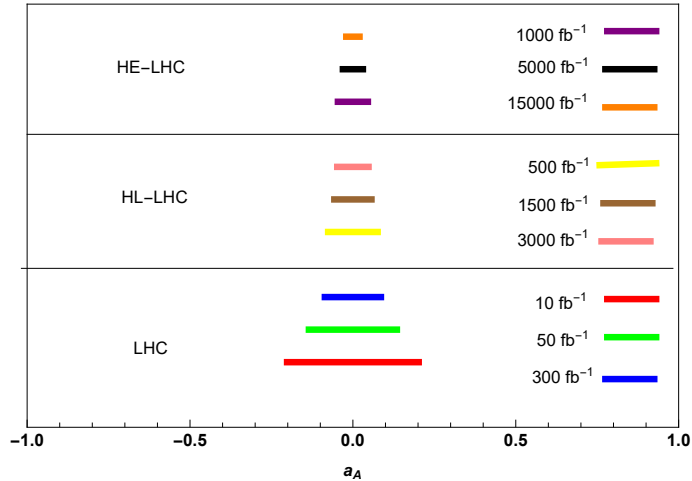}
\caption{$95\%$ C.L. sensitivity limits of the $a_{A}$ coupling for various values of integrated luminosities through the process $pp\rightarrow p \gamma^{*} p \rightarrow p t \bar{t} X$ at the LHC, the HL-LHC and the HE-LHC.
\label{fig7}}
\end{figure}

\begin{figure} [!h]
\includegraphics{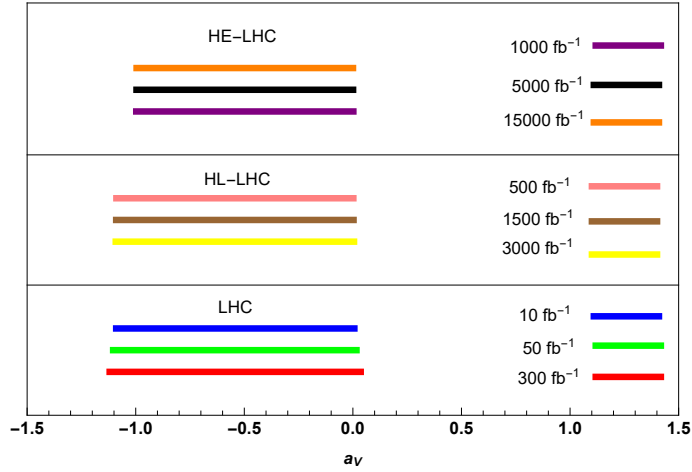}
\caption{Same as in Fig. 7, but for the anomalous $a_{V}$ coupling.
\label{fig8}}
\end{figure}

Similarly, in Figs. 9-11, we present $95\%$ C.L. contours for $a_{A}$ and $a_{V}$ couplings couplings for the process $pp\rightarrow p \gamma p \rightarrow p t \bar{t} X$ at the LHC, the HL-LHC and the HE-LHC for different integrated luminosities. We observe from these figures that the strongest constraint on the anomalous couplings comes from the HE-LHC with 15 ab$^{-1}$.

\begin{figure} [!h]
\includegraphics{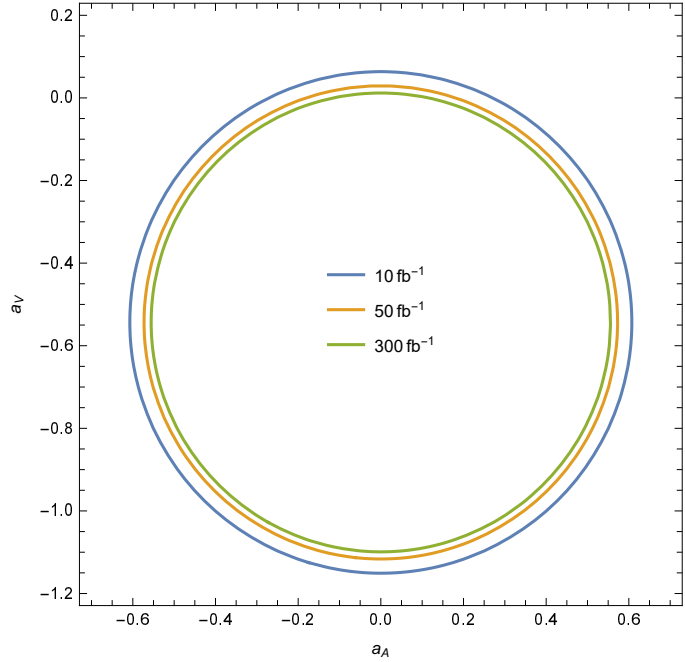}
\caption{For semileptonic channel, contours at $95\%$ C. L. for the anomalous $a_{A}$ and $a_{V}$ couplings for the process $pp\rightarrow p \gamma^{*} p \rightarrow p t \bar{t} X$ at the LHC.
\label{fig9}}
\end{figure}

\begin{figure} [!h]
\includegraphics{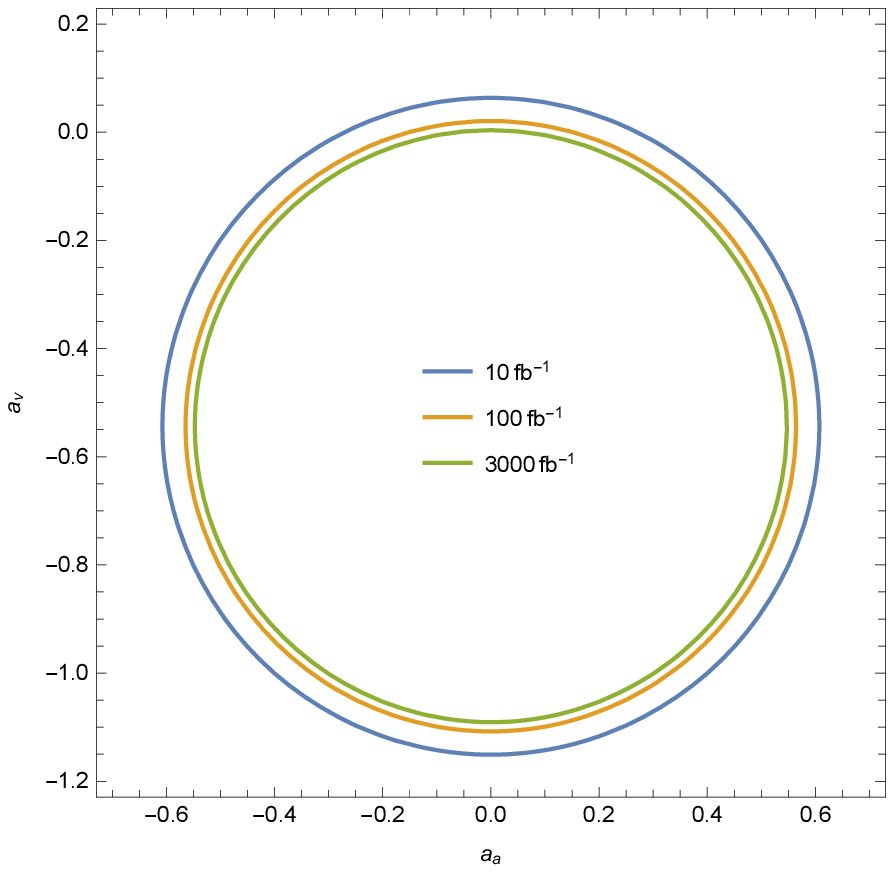}
\caption{Same as in Fig. 9, but for the HL-LHC.
\label{fig10}}
\end{figure}

\begin{figure} [!h]
\includegraphics{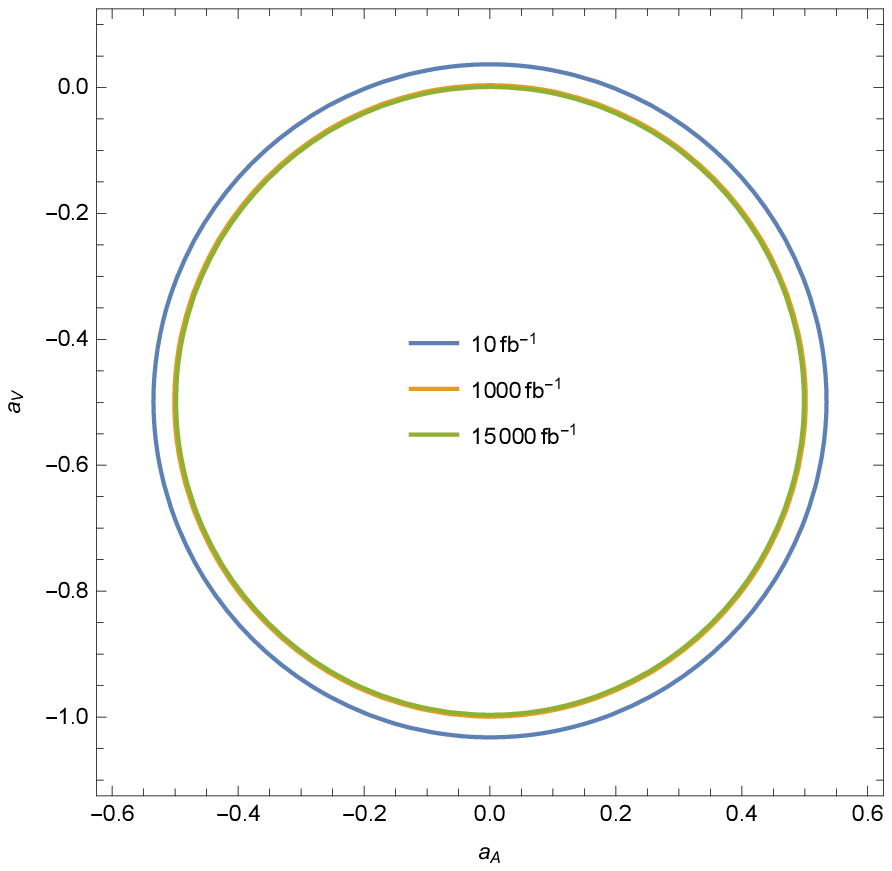}
\caption{Same as in Fig. 9, but for the HE-LHC.
\label{fig11}}
\end{figure}

For pure and semileptonic decay channels, the limits obtained at $68, 90, 95\%$ C.L. on the electromagnetic dipole moments of the top quark via the process $pp\rightarrow p \gamma p \rightarrow p t \bar{t} X$ at the LHC-14 TeV with 300 fb$^{-1}$, the HL-LHC-14 TeV with 3000 fb$^{-1}$ and the HE-LHC-27 TeV with 15000 fb$^{-1}$ are presented in Tables II-XIX.

\begin{table} [!h]
\caption{Limits at $68\%$ Confidence Level on the anomalous $a_{A}$ and $a_{V}$ couplings at the LHC via $t \bar{t}$ production pure leptonic decay channel with integrated luminosities of $10, 30, 50, 100, 200$ and $300$ fb$^{-1}$ for systematic errors of $0,3\%$ and $5\%$.}
\begin{ruledtabular}
\begin{tabular}{cccc}
Luminosity($fb^{-1}$)&$\delta_{sys}$&$|a_{A}|$ & $a_{V}$  \\
\hline
$10$&$0\%$&$0.2022$ &$ (-1.1235;0.0364)$ \\
$10$&$3\%$&$0.2176$ &$ (-1.1290;0.0419)$ \\
$10$&$5\%$&$0.2388$ &$ (-1.1372;0.0501)$ \\
\hline
$30$&$0\%$&$0.1536$ &$ (-1.1084;0.0213)$ \\
$30$&$3\%$&$0.1832$ &$ (-1.1171;0.0300)$ \\
$30$&$5\%$&$0.2150$ &$ (-1.1281;0.0409)$ \\
\hline
$50$&$0\%$&$0.1352$ &$ (-1.1036; 0.0165)$ \\
$50$&$3\%$&$0.1733$ &$ (-1.1141;0.0269)$ \\
$50$&$5\%$&$0.2091$ &$ (-1.1259;0.0388)$ \\
\hline
$100$&$0\%$&$0.1137$ &$ (-1.0988;0.0117)$ \\
$100$&$3\%$&$0.1647$ &$ (-1.1115;0.0244)$ \\
$100$&$5\%$&$0.2044$ &$ (-1.1243;0.0371)$ \\
\hline
$200$&$0\%$&$0.0956$ &$ (-1.0954;0.0083)$ \\
$200$&$3\%$&$0.1598$ &$ (-1.1101;0.0230)$ \\
$200$&$5\%$&$0.1998$ &$ (-1.1234;0.0363)$ \\
\hline
$300$&$0\%$&$0.0864$ &$ (-1.0939;0.0068)$ \\
$300$&$3\%$&$0.1580$ &$ (-1.1096;0.0225)$ \\
$300$&$5\%$&$0.2011$ &$ (-1.1231;0.0360)$ \\
\hline
\end{tabular}
\end{ruledtabular}
\end{table}

\begin{table} [!h]
\caption{Same as in Table II, but for $90\%$ C.L.}
\begin{ruledtabular}
\begin{tabular}{cccc}
Luminosity($fb^{-1}$)&$\delta_{sys}$&$|a_{A}|$ & $a_{V}$  \\
\hline
$10$&$0\%$&$0.2291$ &$ (-1.1334;0.0463)$ \\
$10$&$3\%$&$0.2464$ &$ (-1.1404;0.0532)$ \\
$10$&$5\%$&$0.2705$ &$ (-1.1507;0.0636)$ \\
\hline
$30$&$0\%$&$0.1740$ &$ (-1.1143;0.0271)$ \\
$30$&$3\%$&$0.2075$ &$ (-1.1253;0.0382)$ \\
$30$&$5\%$&$0.2435$ &$ (-1.1392;0.0520)$ \\
\hline
$50$&$0\%$&$0.1532$ &$ (-1.1083;0.0211)$ \\
$50$&$3\%$&$0.1963$ &$ (-1.1215;0.0343)$ \\
$50$&$5\%$&$0.2369$ &$ (-1.1365;0.0494)$ \\
\hline
$100$&$0\%$&$0.1288$ &$ (-1.1021;0.0150)$ \\
$100$&$3\%$&$0.1865$ &$ (-1.1182;0.0311)$ \\
$100$&$5\%$&$0.2316$ &$ (-1.1344;0.0472)$ \\
\hline
$200$&$0\%$&$0.1083$ &$ (-1.0978;0.0106)$ \\
$200$&$3\%$&$0.1810$ &$ (-1.1164;0.0293)$ \\
$200$&$5\%$&$0.2263$ &$ (-1.1333;0.0461)$ \\
\hline
$300$&$0\%$&$0.0978$ &$ (-1.0958;0.0087)$ \\
$300$&$3\%$&$0.1790$ &$ (-1.1158;0.0287)$ \\
$300$&$5\%$&$0.2278$ &$ (-1.1329;0.0458)$ \\
\hline
\end{tabular}
\end{ruledtabular}
\end{table}

\begin{table} [!h]
\caption{Same as in Table II, but for $95\%$ C.L.}
\begin{ruledtabular}
\begin{tabular}{cccc}
Luminosity($fb^{-1}$)&$\delta_{sys}$&$|a_{A}|$ & $a_{V}$  \\
\hline
$10$&$0\%$&$0.2831$ &$ (-1.1564;0.0693)$ \\
$10$&$3\%$&$0.3046$ &$ (-1.1666;0.0795)$ \\
$10$&$5\%$&$0.3343$ &$ (-1.1817;0.0945)$ \\
\hline
$30$&$0\%$&$0.2151$ &$ (-1.1281;0.0410)$ \\
$30$&$3\%$&$0.2564$ &$ (-1.1445;0.0574)$ \\
$30$&$5\%$&$0.3010$ &$ (-1.16491;0.0777)$ \\
\hline
$50$&$0\%$&$0.1893$ &$ (-1.1191;0.0320)$ \\
$50$&$3\%$&$0.2427$ &$ (-1.1388;0.0517)$ \\
$50$&$5\%$&$0.2928$ &$ (-1.1609;0.0738)$ \\
\hline
$100$&$0\%$&$0.1592$ &$ (-1.1099;0.0228)$ \\
$100$&$3\%$&$0.2305$ &$ (-1.1340;0.0468)$ \\
$100$&$5\%$&$0.2862$ &$ (-1.1578;0.0707)$ \\
\hline
$200$&$0\%$&$0.1338$ &$ (-1.1033;0.0162)$ \\
$200$&$3\%$&$0.2237$ &$ (-1.1313;0.0442)$ \\
$200$&$5\%$&$0.2797$ &$ (-1.156;0.0691)$ \\
\hline
$300$&$0\%$&$0.1209$ &$ (-1.1004;0.0133)$ \\
$300$&$3\%$&$0.2213$ &$ (-1.1304;0.0433)$ \\
$300$&$5\%$&$0.2815$ &$ (-1.1557;0.0685)$ \\
\hline
\end{tabular}
\end{ruledtabular}
\end{table}

\begin{table} [!h]
\caption{Limits at $68\%$ Confidence Level on the anomalous $a_{A}$ and $a_{V}$ couplings at the LHC via $t \bar{t}$ production semileptonic decay channel with integrated luminosities of $10, 30, 50, 100, 200$ and $300$ fb$^{-1}$ for systematic errors of $0,3\%$ and $5\%$.}
\begin{ruledtabular}
\begin{tabular}{cccc}
Luminosity($fb^{-1}$)&$\delta_{sys}$&$|a_{A}|$ & $a_{V}$  \\
\hline
$10$&$0\%$&$0.1732$ &$ (-1.11406;0.02694)$ \\
$10$&$3\%$&$0.1958$ &$ (-1.1214;0.0341)$ \\
$10$&$5\%$&$0.2231$ &$ (-1.1387;0.0440)$ \\
\hline
$30$&$0\%$&$0.1316$ &$ (-1.1028;0.0157)$ \\
$30$&$3\%$&$0.1717$ &$ (-1.1072;0.0264)$ \\
$30$&$5\%$&$0.2082$ &$ (-1.1175;0.0385)$ \\
\hline
$50$&$0\%$&$0.1158$ &$ (-1.0993;0.0122)$ \\
$50$&$3\%$&$0.1654$ &$ (-1.1027;0.0246)$ \\
$50$&$5\%$&$0.2048$ &$ (-1.1108;0.0373)$ \\
\hline
$100$&$0\%$&$0.0974$ &$ (-1.0957;0.0086)$ \\
$100$&$3\%$&$0.1602$ &$ (-1.0982;0.0231)$ \\
$100$&$5\%$&$0.2021$ &$ (-1.1039;0.0363)$ \\
\hline
$200$&$0\%$&$0.0819$ &$ (-1.0932;0.0061)$ \\
$200$&$3\%$&$0.1574$ &$ (-1.0949;0.0223)$ \\
$200$&$5\%$&$0.2007$ &$ (-1.0990;0.0359)$ \\
\hline
$300$&$0\%$&$0.0740$ &$ (-1.0921;0.0050)$ \\
$300$&$3\%$&$0.1564$ &$ (-1.0935;0.0220)$ \\
$300$&$5\%$&$0.2003$ &$ (-1.0969;0.0382)$ \\
\hline
\end{tabular}
\end{ruledtabular}
\end{table}

\begin{table} [!h]
\caption{Same as in Table VI, but for $90\%$ C.L.}
\begin{ruledtabular}
\begin{tabular}{cccc}
Luminosity($fb^{-1}$)&$\delta_{sys}$&$|a_{A}|$ & $a_{V}$  \\
\hline
$10$&$0\%$&$0.1962$ &$ (-1.1214;0.0343)$ \\
$10$&$3\%$&$0.2218$ &$ (-1.1306;0.0435)$ \\
$10$&$5\%$&$0.2528$ &$ (-1.1430;0.0559)$ \\
\hline
$30$&$0\%$&$0.1491$ &$ (-1.1072;0.0200)$ \\
$30$&$3\%$&$0.1945$ &$ (-1.1208;0.0337)$ \\
$30$&$5\%$&$0.2358$ &$ (-1.1361;0.0489)$ \\
\hline
$50$&$0\%$&$0.1312$ &$ (-1.1027;0.0156)$ \\
$50$&$3\%$&$0.1874$ &$ (-1.1185;0.0313)$ \\
$50$&$5\%$&$0.232$ &$ (-1.1345;0.0474)$ \\
\hline
$100$&$0\%$&$0.1103$ &$ (-1.0982;0.0110)$ \\
$100$&$3\%$&$0.1814$ &$ (-1.1166;0.0295)$ \\
$100$&$5\%$&$0.2290$ &$ (-1.1333;0.0462)$ \\
\hline
$200$&$0\%$&$0.0928$ &$ (-1.0949;0.0078)$ \\
$200$&$3\%$&$0.1783$ &$ (-1.1156;0.0285)$ \\
$200$&$5\%$&$0.2274$ &$ (-1.1327;0.0456)$ \\
\hline
$300$&$0\%$&$0.0838$ &$ (-1.0935;0.0064)$ \\
$300$&$3\%$&$0.1772$ &$ (-1.1152;0.0281)$ \\
$300$&$5\%$&$0.2269$ &$ (-1.1325;0.0454)$ \\
\hline
\end{tabular}
\end{ruledtabular}
\end{table}

\begin{table} [!h]
\caption{Same as in Table VI, but for $95\%$ C.L.}
\begin{ruledtabular}
\begin{tabular}{cccc}
Luminosity($fb^{-1}$)&$\delta_{sys}$&$|a_{A}|$ & $a_{V}$  \\
\hline
$10$&$0\%$&$0.2425$ &$ (-1.1387;0.0516)$ \\
$10$&$3\%$&$0.2741$ &$ (-1.1523;0.0652)$ \\
$10$&$5\%$&$0.3124$ &$ (-1.1705;0.0834)$ \\
\hline
$30$&$0\%$&$0.1842$ &$ (-1.1175;0.0303)$ \\
$30$&$3\%$&$0.2403$ &$ (-1.1378;0.0507)$ \\
$30$&$5\%$&$0.2915$ &$ (-1.1603;0.0732)$ \\
\hline
$50$&$0\%$&$0.1621$ &$ (-1.1108;0.0236)$ \\
$50$&$3\%$&$0.2315$ &$ (-1.1344;0.0472)$ \\
$50$&$5\%$&$0.2867$ &$ (-1.1581;0.0710)$ \\
\hline
$100$&$0\%$&$0.1363$ &$ (-1.1039;0.0168)$ \\
$100$&$3\%$&$0.2242$ &$ (-1.1315;0.0444)$ \\
$100$&$5\%$&$0.2830$ &$ (-1.1563;0.0692)$ \\
\hline
$200$&$0\%$&$0.1146$ &$ (-1.0990;0.0119)$ \\
$200$&$3\%$&$0.2203$ &$ (-1.1300;0.0429)$ \\
$200$&$5\%$&$0.2810$ &$ (-1.1554;0.0683)$ \\
\hline
$300$&$0\%$&$0.1036$ &$ (-1.0969;0.0098)$ \\
$300$&$3\%$&$0.2189$ &$ (-1.1295;0.0424)$ \\
$300$&$5\%$&$0.2804$ &$ (-1.1552;0.068)$ \\
\hline
\end{tabular}
\end{ruledtabular}
\end{table}

\begin{table} [!h]
\caption{Limits at $68\%$ Confidence Level on the anomalous $a_{A}$ and $a_{V}$ couplings at the HL-LHC via $t \bar{t}$ production pure leptonic decay channel with integrated luminosities of $500, 1000, 1500, 2000, 2500$ and $3000$ fb$^{-1}$ for systematic errors of $0,3\%$ and $5\%$.}
\begin{ruledtabular}
\begin{tabular}{cccc}
Luminosity($fb^{-1}$)&$\delta_{sys}$&$|a_{A}|$ & $a_{V}$  \\
\hline
$500$&$0\%$&$0.0760$ &$ (-1.0924;0.0053)$ \\
$500$&$3\%$&$0.1566$ &$ (-1.1092;0.0221)$ \\
$500$&$5\%$&$0.2004$ &$ (-1.1227;0.0357)$ \\
\hline
$1000$&$0\%$&$0.0639$ &$ (-1.0908;0.0037)$ \\
$1000$&$3\%$&$0.1555$ &$ (-1.1089;0.0218)$ \\
$1000$&$5\%$&$0.1999$ &$ (-1.1227;0.0355)$ \\
\hline
$1500$&$0\%$&$0.0577$ &$ (-1.0901;0.0030)$ \\
$1500$&$3\%$&$0.1551$ &$ (-1.1088;0.0217)$ \\
$1500$&$5\%$&$0.1997$ &$ (-1.1226;0.0355)$ \\
\hline
$2000$&$0\%$&$0.0537$ &$ (-1.0897;0.0026)$ \\
$2000$&$3\%$&$0.1550$ &$ (-1.1087;0.0216)$ \\
$2000$&$5\%$&$0.1995$ &$ (-1.1225;0.0355)$ \\
\hline
$2500$&$0\%$&$0.0508$ &$ (-1.0894;0.0023)$ \\
$2500$&$3\%$&$0.1548$ &$ (-1.1087;0.0216)$ \\
$2500$&$5\%$&$0.1995$ &$ (-1.1225;0.0354)$ \\
\hline
$3000$&$0\%$&$0.0485$ &$ (-1.0892;0.0021)$ \\
$3000$&$3\%$&$0.1548$ &$ (-1.1087;0.0216)$ \\
$3000$&$5\%$&$0.1995$ &$ (-1.1225;0.0354)$ \\
\hline
\end{tabular}
\end{ruledtabular}
\end{table}

\begin{table} [!h]
\caption{Same as in Table VIII, but for $90\%$ C.L.}
\begin{ruledtabular}
\begin{tabular}{cccc}
Luminosity($fb^{-1}$)&$\delta_{sys}$&$|a_{A}|$ & $a_{V}$  \\
\hline
$500$&$0\%$&$0.0861$ &$ (-1.0939;0.0067)$ \\
$500$&$3\%$&$0.1774$ &$ (-1.1153;0.0287)$ \\
$500$&$5\%$&$0.2270$ &$ (-1.1324;0.0455)$ \\
\hline
$1000$&$0\%$&$0.0724$ &$ (-1.0919;0.0048)$ \\
$1000$&$3\%$&$0.1762$ &$ (-1.1149;0.0278)$ \\
$1000$&$5\%$&$0.2264$ &$ (-1.1324;0.0452)$ \\
\hline
$1500$&$0\%$&$0.0654$ &$ (-1.0910;0.0039)$ \\
$1500$&$3\%$&$0.1757$ &$ (-1.1148;0.0277)$ \\
$1500$&$5\%$&$0.2262$ &$ (-1.1323;0.0452)$ \\
\hline
$2000$&$0\%$&$0.0609$ &$ (-1.0905;0.0034)$ \\
$2000$&$3\%$&$0.1755$ &$ (-1.1147;0.0276)$ \\
$2000$&$5\%$&$0.2260$ &$ (-1.1323;0.0451)$ \\
\hline
$2500$&$0\%$&$0.0576$ &$ (-1.0901;0.0030)$ \\
$2500$&$3\%$&$0.1754$ &$ (-1.1147;0.0276)$ \\
$2500$&$5\%$&$0.2260$ &$ (-1.1323;0.0451)$ \\
\hline
$3000$&$0\%$&$0.0550$ &$ (-1.0899;0.0027)$ \\
$3000$&$3\%$&$0.1753$ &$ (-1.1147;0.0275)$ \\
$3000$&$5\%$&$0.2260$ &$ (-1.1323;0.0451)$ \\
\hline
\end{tabular}
\end{ruledtabular}
\end{table}

\begin{table} [!h]
\caption{Same as in Table VIII, but for $95\%$ C.L.}
\begin{ruledtabular}
\begin{tabular}{cccc}
Luminosity($fb^{-1}$)&$\delta_{sys}$&$|a_{A}|$ & $a_{V}$  \\
\hline
$500$&$0\%$&$0.1064$ &$ (-1.0974;0.0103)$ \\
$500$&$3\%$&$0.2193$ &$ (-1.1296;0.0425)$ \\
$500$&$5\%$&$0.2805$ &$ (-1.1549;0.0681)$ \\
\hline
$1000$&$0\%$&$0.0895$ &$ (-1.0944;0.0073)$ \\
$1000$&$3\%$&$0.2177$ &$ (-1.1291;0.0420)$ \\
$1000$&$5\%$&$0.2798$ &$ (-1.1549;0.0678)$ \\
\hline
$1500$&$0\%$&$0.0809$ &$ (-1.0931;0.0059)$ \\
$1500$&$3\%$&$0.2172$ &$ (-1.1289;0.0418)$ \\
$1500$&$5\%$&$0.2796$ &$ (-1.1548;0.0677)$ \\
\hline
$2000$&$0\%$&$0.0752$ &$ (-1.0923;0.0051)$ \\
$2000$&$3\%$&$0.2169$ &$ (-1.1288;0.0417)$ \\
$2000$&$5\%$&$0.2794$ &$ (-1.1548;0.0676)$ \\
\hline
$2500$&$0\%$&$0.0712$ &$ (-1.0917;0.0046)$ \\
$2500$&$3\%$&$0.2168$ &$ (-1.1287;0.0416)$ \\
$2500$&$5\%$&$0.2794$ &$ (-1.1548;0.0676)$ \\
\hline
$3000$&$0\%$&$0.0680$ &$ (-1.0913;0.0042)$ \\
$3000$&$3\%$&$0.2167$ &$ (-1.1287;0.0416)$ \\
$3000$&$5\%$&$0.2794$ &$ (-1.15482;0.0675)$ \\
\hline
\end{tabular}
\end{ruledtabular}
\end{table}

\begin{table} [!h]
\caption{Limits at $68\%$ Confidence Level on the anomalous $a_{A}$ and $a_{V}$ couplings at the HL-LHC via $t \bar{t}$ production semileptonic decay channel with integrated luminosities of $500, 1000, 1500, 2000, 2500$ and $3000$ fb$^{-1}$ for systematic errors of $0,3\%$ and $5\%$.}
\begin{ruledtabular}
\begin{tabular}{cccc}
Luminosity($fb^{-1}$)&$\delta_{sys}$&$|a_{A}|$ & $a_{V}$  \\
\hline
$500$&$0\%$&$0.0651$ &$ (-1.0910;0.0038)$ \\
$500$&$3\%$&$ 0.1556$ &$ (-1.1089;0.0220)$ \\
$500$&$5\%$&$0.1999$ &$ (-1.1227;0.0357)$ \\
\hline
$1000$&$0\%$&$0.0547$ &$ (-1.0898;0.0027)$ \\
$1000$&$3\%$&$0.1550$ &$ (-1.1088;0.0218)$ \\
$1000$&$5\%$&$0.1996$ &$ (-1.1226;0.0356)$ \\
\hline
$1500$&$0\%$&$0.0495$ &$ (-1.0893;0.0022)$ \\
$1500$&$3\%$&$0.1548$ &$ (-1.1087;0.0216)$ \\
$1500$&$5\%$&$0.1996$ &$ (-1.1226;0.0355)$ \\
\hline
$2000$&$0\%$&$0.0460$ &$ (-1.0890;0.0019)$ \\
$2000$&$3\%$&$0.1547$ &$ (-1.1086;0.0216)$ \\
$2000$&$5\%$&$0.1995$ &$ (-1.1225;0.0354)$ \\
\hline
$2500$&$0\%$&$0.0435$ &$ (-1.0888;0.0017)$ \\
$2500$&$3\%$&$0.1546$ &$ (-1.1086;0.0215)$ \\
$2500$&$5\%$&$0.1995$ &$ (-1.1225;0.0354)$ \\
\hline
$3000$&$0\%$&$0.0416$ &$ (-1.0887;0.0015)$ \\
$3000$&$3\%$&$0.1546$ &$ (-1.1086;0.0215)$ \\
$3000$&$5\%$&$0.1994$ &$ (-1.1225;0.0354)$ \\
\hline
\end{tabular}
\end{ruledtabular}
\end{table}

\begin{table} [!h]
\caption{Same as in Table XI, but for $90\%$ C.L.}
\begin{ruledtabular}
\begin{tabular}{cccc}
Luminosity($fb^{-1}$)&$\delta_{sys}$&$|a_{A}|$ & $a_{V}$  \\
\hline
$500$&$0\%$&$0.0738$ &$ (-1.0921;0.0050)$ \\
$500$&$3\%$&$0.1763$ &$ (-1.1150;0.0280)$ \\
$500$&$5\%$&$0.2264$ &$ (-1.1324;0.0454)$ \\
\hline
$1000$&$0\%$&$0.0620$ &$ (-1.0906;0.0035)$ \\
$1000$&$3\%$&$0.1756$ &$ (-1.1147;0.0278)$ \\
$1000$&$5\%$&$0.2261$ &$ (-1.1322;0.0453)$ \\
\hline
$1500$&$0\%$&$0.0560$ &$ (-1.0900;0.0029)$ \\
$1500$&$3\%$&$0.1754$ &$ (-1.1147;0.0276)$ \\
$1500$&$5\%$&$0.2260$ &$ (-1.1322;0.0452)$ \\
\hline
$2000$&$0\%$&$0.0521$ &$ (-1.0896;0.0025)$ \\
$2000$&$3\%$&$0.1752$ &$ (-1.1146;0.0275)$ \\
$2000$&$5\%$&$0.2260$ &$ (-1.1322;0.0451)$ \\
\hline
$2500$&$0\%$&$0.0493$ &$ (-1.0893;0.0022)$ \\
$2500$&$3\%$&$0.1752$ &$ (-1.1146;0.0275)$ \\
$2500$&$5\%$&$0.2259$ &$ (-1.1322;0.0450)$ \\
\hline
$3000$&$0\%$&$0.0471$ &$ (-1.0891;0.0020)$ \\
$3000$&$3\%$&$0.1751$ &$ (-1.1146;0.0275)$ \\
$3000$&$5\%$&$0.2259$ &$ (-1.1322;0.0450)$ \\
\hline
\end{tabular}
\end{ruledtabular}
\end{table}

\begin{table} [!h]
\caption{Same as in Table XI, but for $95\%$ C.L.}
\begin{ruledtabular}
\begin{tabular}{cccc}
Luminosity($fb^{-1}$)&$\delta_{sys}$&$|a_{A}|$ & $a_{V}$  \\
\hline
$500$&$0\%$&$0.0912$ &$ (-1.0947;0.0076)$ \\
$500$&$3\%$&$0.2178$ &$ (-1.1291;0.0424)$ \\
$500$&$5\%$&$0.2799$ &$ (-1.1549;0.0679)$ \\
\hline
$1000$&$0\%$&$0.0767$ &$ (-1.0925;0.0054)$ \\
$1000$&$3\%$&$0.2170$ &$ (-1.1288;0.0420)$ \\
$1000$&$5\%$&$0.2795$ &$ (-1.1547;0.0678)$ \\
\hline
$1500$&$0\%$&$0.0693$ &$ (-1.0915;0.0044)$ \\
$1500$&$3\%$&$0.2167$ &$ (-1.1287;0.0417)$ \\
$1500$&$5\%$&$0.2793$ &$ (-1.1547;0.0676)$ \\
\hline
$2000$&$0\%$&$0.0644$ &$ (-1.0909;0.0038)$ \\
$2000$&$3\%$&$0.2166$ &$ (-1.1286;0.0415)$ \\
$2000$&$5\%$&$0.2793$ &$ (-1.1546;0.0675)$ \\
\hline
$2500$&$0\%$&$0.0609$ &$ (-1.0905;0.0034)$ \\
$2500$&$3\%$&$0.2165$ &$ (-1.1286;0.0415)$ \\
$2500$&$5\%$&$0.2792$ &$ (-1.1546;0.0675)$ \\
\hline
$3000$&$0\%$&$0.0582$ &$ (-1.0902;0.0031)$ \\
$3000$&$3\%$&$0.2164$ &$ (-1.1286;0.0415)$ \\
$3000$&$5\%$&$0.2792$ &$ (-1.1546;0.0675)$ \\
\hline
\end{tabular}
\end{ruledtabular}
\end{table}

\begin{table} [!h]
\caption{Limits at $68\%$ Confidence Level on the anomalous $a_{A}$ and $a_{V}$ couplings at the HE-LHC via $t \bar{t}$ production pure leptonic decay channel with integrated luminosities of $1000, 3000, 5000, 10000$ and $15000$ fb$^{-1}$ for systematic errors of $0,3\%$ and $5\%$.}
\begin{ruledtabular}
\begin{tabular}{cccc}
Luminosity($fb^{-1}$)&$\delta_{sys}$&$|a_{A}|$ & $a_{V}$  \\
\hline
$1000$&$0\%$&$0.0460$ &$ (-1.1546;0.0021)$ \\
$1000$&$3\%$&$0.1488$ &$ (-1.0173;0.0217)$ \\
$1000$&$5\%$&$0.1918$ &$ (-1.0312;0.0356)$ \\
\hline
$3000$&$0\%$&$0.0349$ &$ (-0.9967;0.0012)$ \\
$3000$&$3\%$&$0.1487$ &$ (-1.0172;0.0216)$ \\
$3000$&$5\%$&$0.1917$ &$ (-1.0311;0.0356)$ \\
\hline
$5000$&$0\%$&$0.0307$ &$ (-0.9964;0.0009)$ \\
$5000$&$3\%$&$0.1486$ &$ (-1.0172;0.0216)$ \\
$5000$&$5\%$&$0.1917$ &$ (-1.0311;0.0356)$ \\
\hline
$10000$&$0\%$&$0.0258$ &$ (-0.9962;0.0006)$ \\
$10000$&$3\%$&$0.1485$ &$ (-1.0172;0.0216)$ \\
$10000$&$5\%$&$0.1917$ &$ (-1.0311;0.0356)$ \\
\hline
$15000$&$0\%$&$0.0233$ &$ (-0.9960;0.0005)$ \\
$15000$&$3\%$&$0.1485$ &$ (-1.0172;0.0216)$ \\
$15000$&$5\%$&$0.1917$ &$ (-1.0311;0.0356)$ \\
\hline
\end{tabular}
\end{ruledtabular}
\end{table}

\begin{table} [!h]
\caption{Same as in Table XIV, but for $90\%$ C.L.}
\begin{ruledtabular}
\begin{tabular}{cccc}
Luminosity($fb^{-1}$)&$\delta_{sys}$&$|a_{A}|$ & $a_{V}$  \\
\hline
$1000$&$0\%$&$0.0521$ &$ (-0.9982;0.0027)$ \\
$1000$&$3\%$&$0.1685$ &$ (-1.0233;0.0277)$ \\
$1000$&$5\%$&$0.2173$ &$ (-1.0409;0.0453)$ \\
\hline
$3000$&$0\%$&$0.0396$ &$ (-0.9971;0.0015)$ \\
$3000$&$3\%$&$0.1684$ &$ (-1.0232;0.0273)$ \\
$3000$&$5\%$&$0.2172$ &$ (-1.0408;0.0452)$ \\
\hline
$5000$&$0\%$&$0.0348$ &$ (-0.9967;0.0012)$ \\
$5000$&$3\%$&$0.1683$ &$ (-1.0232;0.0276)$ \\
$5000$&$5\%$&$0.2171$ &$ (-1.0408;0.0452)$ \\
\hline
$10000$&$0\%$&$0.0293$ &$ (-0.9964;0.0008)$ \\
$10000$&$3\%$&$0.1682$ &$ (-1.0232;0.0276)$ \\
$10000$&$5\%$&$0.2171$ &$ (-1.0408;0.0452)$ \\
\hline
$15000$&$0\%$&$0.0264$ &$ (-0.9962;0.0007)$ \\
$15000$&$3\%$&$0.1682$ &$ (-1.0231;0.0276)$ \\
$15000$&$5\%$&$0.2171$ &$ (-1.0408;0.0452)$ \\
\hline
\end{tabular}
\end{ruledtabular}
\end{table}

\begin{table} [!h]
\caption{Same as in Table XIV, but for $95\%$ C.L.}
\begin{ruledtabular}
\begin{tabular}{cccc}
Luminosity($fb^{-1}$)&$\delta_{sys}$&$|a_{A}|$ & $a_{V}$  \\
\hline
$1000$&$0\%$&$0.0644$ &$ (-0.9996;0.0041)$ \\
$1000$&$3\%$&$0.2083$ &$ (-1.0373;0.0418)$ \\
$1000$&$5\%$&$0.2685$ &$ (-1.0633;0.0678)$ \\
\hline
$3000$&$0\%$&$0.0489$ &$ (-0.9979;0.0024)$ \\
$3000$&$3\%$&$0.2081$ &$ (-1.0372;0.0418)$ \\
$3000$&$5\%$&$0.2684$ &$ (-1.0632;0.0677)$ \\
\hline
$5000$&$0\%$&$0.0430$ &$ (-0.9973;0.0018)$ \\
$5000$&$3\%$&$0.2080$ &$ (-1.0372;0.0416)$ \\
$5000$&$5\%$&$0.2683$ &$ (-1.0632;0.0677)$ \\
\hline
$10000$&$0\%$&$0.0362$ &$ (-0.9968;0.0013)$ \\
$10000$&$3\%$&$0.2079$ &$ (-1.0372;0.0416)$ \\
$10000$&$5\%$&$0.2683$ &$ (-1.0632;0.0677)$ \\
\hline
$15000$&$0\%$&$0.0327$ &$ (-0.9966;0.0010)$ \\
$15000$&$3\%$&$0.2079$ &$ (-1.0372;0.0416)$ \\
$15000$&$5\%$&$0.2683$ &$ (-1.0632;0.0677)$ \\
\hline
\end{tabular}
\end{ruledtabular}
\end{table}

\begin{table} [!h]
\caption{Limits at $68\%$ Confidence Level on the anomalous $a_{A}$ and $a_{V}$ couplings at the HE-LHC via $t \bar{t}$ production semileptonic decay channel with integrated luminosities of $1000, 3000, 5000, 10000$ and $15000$ fb$^{-1}$ for systematic errors of $0,3\%$ and $5\%$.}
\begin{ruledtabular}
\begin{tabular}{cccc}
Luminosity($fb^{-1}$)&$\delta_{sys}$&$|a_{A}|$ & $a_{V}$  \\
\hline
$1000$&$0\%$&$0.0394$ &$ (-0.9970;0.0015)$ \\
$1000$&$3\%$&$0.1486$ &$ (-1.0172;0.0217)$ \\
$1000$&$5\%$&$0.1917$ &$ (-1.0312;0.0356)$ \\
\hline
$3000$&$0\%$&$0.0299$ &$ (-0.9964;0.0009)$ \\
$3000$&$3\%$&$0.1485$ &$ (-1.0172;0.0216)$ \\
$3000$&$5\%$&$0.1917$ &$ (-1.0311;0.0356)$ \\
\hline
$5000$&$0\%$&$0.0263$ &$ (-0.9962;0.0007)$ \\
$5000$&$3\%$&$0.1485$ &$ (-1.0172;0.0216)$ \\
$5000$&$5\%$&$0.1917$ &$ (-1.0311;0.0356)$ \\
\hline
$10000$&$0\%$&$0.0221$ &$ (-0.9960;0.0004)$ \\
$10000$&$3\%$&$0.1485$ &$ (-1.0172;0.0216)$ \\
$10000$&$5\%$&$0.1917$ &$ (-1.0311;0.0356)$ \\
\hline
$15000$&$0\%$&$0.0200$ &$ (-0.9959;0.0003)$ \\
$15000$&$3\%$&$0.1485$ &$ (-1.0172;0.0216)$ \\
$15000$&$5\%$&$0.1917$ &$ (-1.0311;0.0356)$ \\
\hline
\end{tabular}
\end{ruledtabular}
\end{table}

\begin{table} [!h]
\caption{Same as in Table XVII, but for $90\%$ C.L.}
\begin{ruledtabular}
\begin{tabular}{cccc}
Luminosity($fb^{-1}$)&$\delta_{sys}$&$|a_{A}|$ & $a_{V}$  \\
\hline
$1000$&$0\%$&$0.0446$ &$ (-0.9975;0.0020)$ \\
$1000$&$3\%$&$0.1684$ &$ (-1.0232;0.0277)$ \\
$1000$&$5\%$&$0.2172$ &$ (-1.0408;0.0453)$ \\
\hline
$3000$&$0\%$&$0.0339$ &$ (-0.9966;0.0011)$ \\
$3000$&$3\%$&$0.1684$ &$ (-1.0232;0.0276)$ \\
$3000$&$5\%$&$0.2171$ &$ (-1.0408;0.0452)$ \\
\hline
$5000$&$0\%$&$0.0298$ &$ (-0.9964;0.0008)$ \\
$5000$&$3\%$&$0.1682$ &$ (-1.0232;0.0276)$ \\
$5000$&$5\%$&$0.2171$ &$ (-1.0408;0.0452)$ \\
\hline
$10000$&$0\%$&$0.0251$ &$ (-0.9961;0.0006)$ \\
$10000$&$3\%$&$0.1682$ &$ (-1.0232;0.0276)$ \\
$10000$&$5\%$&$0.2171$ &$ (-1.0408;0.0452)$ \\
\hline
$15000$&$0\%$&$0.0226$ &$ (-0.9960;0.0005)$ \\
$15000$&$3\%$&$0.1682$ &$ (-1.0232;0.0276)$ \\
$15000$&$5\%$&$0.2171$ &$ (-1.0408;0.0452)$ \\
\hline
\end{tabular}
\end{ruledtabular}
\end{table}

\begin{table} [!h]
\caption{Same as in Table XVII, but for $95\%$ C.L.}
\begin{ruledtabular}
\begin{tabular}{cccc}
Luminosity($fb^{-1}$)&$\delta_{sys}$&$|a_{A}|$ & $a_{V}$  \\
\hline
$1000$&$0\%$&$0.0551$ &$ (-0.9985;0.0030)$ \\
$1000$&$3\%$&$0.2081$ &$ (-1.0372;0.0418)$ \\
$1000$&$5\%$&$0.2684$ &$ (-1.0632;0.0678)$ \\
\hline
$3000$&$0\%$&$0.0419$ &$ (-0.9973;0.0017)$ \\
$3000$&$3\%$&$0.2079$ &$ (-1.0372;0.0418)$ \\
$3000$&$5\%$&$0.2683$ &$ (-1.0632;0.0677)$ \\
\hline
$5000$&$0\%$&$0.0368$ &$ (-0.9969;0.0013)$ \\
$5000$&$3\%$&$0.2079$ &$ (-1.0372;0.0416)$ \\
$5000$&$5\%$&$0.2683$ &$ (-1.0632;0.0677)$ \\
\hline
$10000$&$0\%$&$0.0310$ &$ (-0.9965;0.0009)$ \\
$10000$&$3\%$&$0.2078$ &$ (-1.0372;0.0416)$ \\
$10000$&$5\%$&$0.2683$ &$ (-1.0632;0.0677)$ \\
\hline
$15000$&$0\%$&$0.0280$ &$ (-0.9963;0.0007)$ \\
$15000$&$3\%$&$0.2078$ &$ (-1.0372;0.0416)$ \\
$15000$&$5\%$&$0.2683$ &$ (-1.0632;0.0677)$ \\
\hline
\end{tabular}
\end{ruledtabular}
\end{table}

In Table II, the best limits obtained on the anomalous $a_{A}$ and $a_{V}$ couplings are $|a_{A}|=0.0864$ and $-1.0939<a_{V}<0.0068$. We observe that the anomalous $a_{A}$ couplings we found from our process for pure leptonic decay channel with $14$ TeV and 300 fb$^{-1}$ are better than those reported in Refs. \cite{6},\cite{8},\cite{13},\cite{14},\cite{15},\cite{16}. However, we compare our results with the limits of Ref. \cite{9}, in which the best limits on $a_{A}$ and $a_{V}$ couplings by probing the process $pp \to p\gamma^*\gamma^*p\to pt\bar t p $ at LHC-33 TeV with $L_{int}=3000$ fb$^{-1}$ are found. We see from Table II  that our limit obtained on $a_{A}$ coupling is nearly the same with those reported in the Ref. \cite{9}.  While the negative part of $a_{V}$ coupling is 2.5 times worse than the limit calculated in Ref. \cite{9}, the positive part of this coupling is 2.5 times better. All Tables show that our best results are given in Table XVII. These are $|a_{A}|=0.0200$ and $-0.9959<a_{V}<0.0003$.  Our result for $a_{A}$ coupling is the same as the result of Ref. \cite{10} which obtains the best limit on the anomalous $a_{A}$ coupling in the literature.

In Table V, the best sensitivities derived  from the process $pp\rightarrow p \gamma^{*} p \rightarrow p t \bar{t} X$ at the LHC with $L_{int}=300$ fb$^{-1}$ are obtained as $|a_{A}|=0.0740$ and $a_{V}=[-1.0921;0.0050]$. As shown in Table XI,  the best sensitivities on $a_{A}$ and $a_{V}$ couplings are $0.0416$ and $[1.0887; 0.0015]$, respectively. However, one can see from Table XVII that the sensitivities on the anomalous couplings are calculated as $|a_{A}|=0.0200$ and $a_{V}=[-0.9959;0.0003]$. We have seen from these Tables that limits on the anomalous $a_{A}$ and $a_{V}$ couplings are improved for increasing integrated luminosities and center-of-mass energies.

We examine the effects on the limits of systematic errors. The best limits obtained by $0\%$ systematic error for $a_{A}$ coupling are almost an order of magnitude better than the results of $5\%$ systematic error. In addition, we find that while the sensitivity obtained on the positive part of $a_{V}$ coupling with $0\%$ systematic error can set more stringent sensitive by three orders of magnitude with respect to the our best sensitivity derived with $5\%$ systematic error, the results obtained on the negative part of $a_{V}$ coupling are nearly the same in both errors. The reason for these behaviors can be easily understood from Figs. 3 and 4. As seen from these figures, while the positive part of $a_{V}$ coupling and $a_{A}$ coupling are strongly dependent on the total cross section, the negative part of $a_{V}$ coupling is very low in relation to the total cross section.

We understand all Tables that the obtained results for the anomalous couplings in the leptonic decay channel are weaker by up to a factor of 0.85 than those related to the hadronic decay channel.

\section{Conclusions}

Investigation of the top quark coupling to photons offers one of the important alternatives to explore new physics beyond the SM such as the electric and magnetic dipole moments of the top quark. However, the electric dipole moment of the top quark is especially interesting since it is very sensitive to possible new sources of CP violation in the lepton and quark sectors.

$\gamma^{*} \gamma^{*}$ and $\gamma^{*} p$ collisions at the LHC provide a suitable platform to examine new physics beyond the SM. $\gamma^{*} p$ collisions have high center-of-mass energy and high luminosity compared to $\gamma^{*} \gamma^{*}$ collision. Furthermore, $\gamma^{*} p$ collisions due to the remnants of only one of the proton beams provide fewer backgrounds according to usual $pp$ deep inelastic scattering.  Moreover, since it has cleaner background, $\gamma^{*}p$ collisions may provide a good opportunity to examine the anomalous $t\bar{t}\gamma$ couplings that define the electromagnetic dipole moments of the top quark.

The anomalous $t\bar{t}\gamma$ coupling has very strong energy dependence due to contributions arising from dimension-six effective operators. Therefore, the total cross section with the anomalous $t\bar{t}\gamma$ coupling has higher energy dependence than the cross section of the SM. In this respect, investigation of the anomalous $t\bar{t}\gamma$ coupling in particle colliders with high center-of-mass energy can be extremely important in determining a possible new physics signal beyond the SM.

For these reasons, we have examined a phenomenological investigation to analyze the sensitivity of the LHC to the anomalous $t\bar{t}\gamma$ vertex taking into account the pure leptonic and the semileptonic decay channels of the top quark pair production in the final states of the process $pp\rightarrow p \gamma p \rightarrow p t \bar{t} X$ at the center-of-mass energies of 14, 27 TeV. Our results show that with a center-of-mass energy of HE-LHC-27 TeV, integrated luminosity of 15 ab$^{-1}$ with the semileptonic decay channel, it is likely that while the LHC can be obtained limits on the top quark's electric dipole moment up to a sensitivity of the order 10$^{-4}$, the limits on the magnetic dipole moment can reach up to a sensitivity of the order 10$^{-2}$.

As a result, $\gamma^{*} p$ collisions at the LHC have a great potential to study the electric and magnetic dipole moments of the top quark.

\pagebreak

\newpage


\begin{thebibliography}{99}

\bibitem{1} S. Chatrchyan \textit{et al}., CMS Collaboration, Phys. Lett. B 716, 30 (2012).
\bibitem{2} G. Aad \textit{et al}., ATLAS Collaboration, Phys. Lett. B 716, 1 (2012).
\bibitem{3} M. Tanabashi \textit{et al}., [Particle Data Group], Phys. Rev. D 98, no. 3, 030001 (2018).
\bibitem{4} D. Atwood, S. Bar-Shalom, G. Eilam and A. Soni, Phys. Rept. 347, 1 (2001).
\bibitem{5} A. Brandenburg and J. P. Ma, Phys. Lett. B 298, 211 (1993).
\bibitem{6} U. Baur, A. Juste, L. H. Orr and D. Rainwater, Phys. Rev. D 71, 054013 (2005).
\bibitem{7} S. M. Etesami, S. Khatibi and M. M. Najafabadi, Eur. Phys. J. C 76, 533 (2016).
\bibitem{8} M. Fael and T. Gehrmann, Phys. Rev. D 88, 033003 (2013).
\bibitem{9}  Sh. Fayazbakhsh, S. Taheri Monfared and M. Mohammadi Najafabadi, Phys. Rev. D 92,
014006 (2015).
\bibitem{10} J. A. Aguilar-Saavedra \textit{et al}., [ECFA/DESY LC Physics Working Group Collaboration], DESY 2001-011, ECFA 2001-209
arXiv:0106315.
\bibitem{11} M. Koksal, A. A. Billur and A. Gutierrez-Rodriguez, Adv. High Energy Phys. 2017, 6738409
(2017).
\bibitem{12} A. A. Billur, M. Koksal and A. Gutierrez-Rodriguez, Phys. Rev. D 96, 056007 (2017).
\bibitem{13} M. Koksal, A. A. Billur, A. Gutierrez-Rodriguez, M. A. Hernandez-Ruiz, arXiv:1905.02564.
\bibitem{14} A. O. Bouzas and F. Larios, Phys. Rev. D 87, 074015 (2013).
\bibitem{15} A. A. Billur, M. Koksal, A. Gutierrez-Rodriguez and M. A. Hernandez-Ruiz, arXiv:1811.10462.
\bibitem{16} A. O. Bouzas and F. Larios, Phys. Rev. D 87, no. 7, 074015 (2013).
\bibitem{epa} V. M. Budnev, I. F. Ginzburg, G. V. Meledin and V. G. Serbo, Phys. Rep. 15, 181 (1975).
\bibitem{epa1} G. Baur \textit{et al}., Phys. Rep. 364, 359 (2002).
\bibitem{epa2} K. Piotrzkowski, Phys. Rev. D 63, 071502 (2001).
\bibitem{s1} I. Sahin and M. Koksal, JHEP 11, 100 (2011).
\bibitem{s2} I. Sahin and A. A. Billur, Phys. Rev. D 83, 035011 (2011).
\bibitem{s3} S. C. Inan and A. V. Kisselev, arXiv::1902.08615.
\bibitem{s4} S. C. Inan and A. V. Kisselev, Eur. Phys. J. C 78, no.9, 729 (2018).
\bibitem{s5} M. Koksal \textit{et al}., Phys. Lett. B 783, 375-380 (2018).
\bibitem{s6} A. A. Billur, Europhys. Lett. 101, 21001 (2013).
\bibitem{s7} M. Koksal and S. C. Inan, Adv. High Energy Phys. 2014, 935840 (2014).
\bibitem{s8} M. Koksal and S. C. Inan, Adv. High Energy Phys. 2014, 315826 (2014).
\bibitem{s9} S. C. Inan, Nucl. Phys. B 897, 289(2015).
\bibitem{s10} P. Xue-An \textit{et al}., Phys. Rev. D 99, 014029 (2019).
\bibitem{s11} A. Senol and M. Koksal, Phys. Lett. B 742, 143 (2015).
\bibitem{s12} D. Alva, T. Han and R. Ruiz, JHEP 1502, 072 (2015).
\bibitem{s13} S. C. Inan and A. A. Billur, Phys.Rev. D 84, 095002 (2011).
\bibitem{s14} B. Sahin and A. A. Billur, Phys.Rev. D 85, 074026 (2012).
\bibitem{s15} B Sahin, Mod.Phys.Lett. A 32, no.37, 1750205 (2017).
\bibitem{s16} B Sahin, Adv. High Energy Phys. 2015, 590397 (2015).
\bibitem{s17} B Sahin and A. A. Billur, Phys.Rev. D 86, 074026 (2012).
\bibitem{s18} A. Esmaili, S. Khatibi and M. M. Najafabadi, Phys. Rev. D 96, 015027 (2017).
\bibitem{s19} A. Senol, Phys. Rev. D 87 073003 (2013).
\bibitem{s20} C. Baldenegro, S. Fichet, G. von Gersdorff, C. Royon,  JHEP 1706, 142 (2017).
\bibitem{s21} C. Baldenegro, S. Fichet, G. von Gersdorff, C. Royon, JHEP 1806, 131 (2018).
\bibitem{s22} S. Fichet, G. V. Gersdorff, B. Lenzi, C. Royon, M. Saimpert, JHEP 1502, 165 (2015).
\bibitem{sa1} G. Akkaya Selcin and \.{I}. \c{S}ahin, Chin. J. Phys. 55, 2305-2317 (2017).
\bibitem{sa2} I. Sahin \textit{et al}., Phys. Rev. D 91, 035017 (2015).
\bibitem{sa3} I. Sahin \textit{et al}., Phys. Rev. D 88,  095016 (2013).
\bibitem{sa4} I. Sahin and B. \c{S}ahin,  Phys.Rev. D 86, 115001 (2012).
\bibitem{sa5} I. Sahin, Phys. Rev. D 85, 033002 (2012).
\bibitem{sa6} H. Sun \textit{et al}., JHEP 1502, 064 (2015).
\bibitem{sa7} H. Sun, Phys. Rev. D 90, 035018 (2014).
\bibitem{sa8} H. Sun, Nucl.Phys. B 886, 691-711 (2014).
\bibitem{sa9} R. Goldouzian and B. Clerbaux, Phys. Rev. D 95, 054014 (2017).
\bibitem{sa10} A. Senol, Int. J. Mod. Phys. A 29, 1450148 (2014).
\bibitem{sa11} A. Senol \textit{et al}., Mod. Phys. Lett. A 29  no.36, 1450186.
\bibitem{Kamenik} J. F. Kamenik, M. Papucci and A. Weiler, {\it Phys. Rev.} {\bf D85}, 071501 (2012).
\bibitem{Baur} U. Baur, A. Juste, L. H. Orr and D. Rainwater, {\it Phys. Rev.} {\bf D71}, 054013 (2005).
\bibitem{Aguilar} J. A. Aguilar-Saavedra, Nucl. Phys. B 812, 181 (2009).
\bibitem{Aguilar1} J. A. Aguilar-Saavedra, M. C. N. Fiolhais and A. Onofre, JHEP {\bf 07}, 180 (2012).
\bibitem{gutt} K. Passek-Kumericki, G. Peters, Phys. Rev. D 78, 033009 (2008).
\bibitem{bud} V. M. Budnev, I. F. Ginzburg, G. V. Meledin and V. G. Serbo, Phys. Rep. 15, 181 (1975).
\bibitem{top} ATLAS Collaboration, Phys. Lett. B 761, 136 (2016).
\end{thebibliography}
\end{document}